\documentclass[journal]{IEEEtran}
\usepackage[english]{babel}
\usepackage[noadjust]{cite}
\usepackage[linesnumbered,ruled,vlined]{algorithm2e}
\ifCLASSINFOpdf
\else
   \usepackage[dvips]{graphicx}
\fi
\usepackage{url}

\usepackage{graphicx}
\usepackage{optidef}
\usepackage{amsmath}
\usepackage{bm}
\usepackage{amssymb}
\usepackage{xcolor}
\usepackage{caption}
\usepackage{subcaption}
\usepackage{lipsum}
\usepackage{siunitx}

\usepackage{titlesec}
\titlespacing\section{0pt}{*0.8}{*0.8}
\titlespacing\subsection{0pt}{*0.7}{*0.7}
\titlespacing\subsubsection{0pt}{*0.6}{*0.6}
\DeclareSIUnit{\belmilliwatt}{Bm}
\DeclareSIUnit{\dBm}{\deci\belmilliwatt}

\begin{document}
\title{ Fusing Channel and Sensor Measurements for Enhancing Predictive Beamforming in UAV-Assisted Massive MIMO Communications}



\author{
Byunghyun Lee, \IEEEmembership{Graduate Student Member, IEEE}, Andrew C. Marcum, \IEEEmembership{Member, IEEE}, David J. Love, \IEEEmembership{Fellow, IEEE}, and James V. Krogmeier, \IEEEmembership{Senior Member, IEEE} 

\vspace{-4mm}

\setlength{\skip\footins}{0.1pt}\thanks{
This work is supported by the National Science Foundation (NSF) under NSF Cooperative Agreement Number EEC-1941529.
}
\setlength{\skip\footins}{1pt}\thanks{Byunghyun Lee, David J. Love, and James V. Krogmeier are with the Department of Electrical and Computer
Engineering, Purdue University, West Lafayette, IN 47907 USA (e-mails:
\{lee4093,djlove,jvk\}@purdue.edu).}
\setlength{\skip\footins}{3pt}\thanks{Andrew C. Marcum is with Raytheon BBN Technologies, Cambridge,
MA 02138 USA (e-mail: andrew.marcum@raytheon.com).}
}

\vspace{-2mm}

\maketitle

\begin{abstract}



Cellular-connected unmanned aerial vehicles (UAVs) represent a promising technology for extending the coverage of 5G and 6G networks in a cost-effective manner.
Additionally, Massive multiple-input multiple-output (MIMO) serves as an effective solution to interference mitigation in cellular-connected UAV communications.
In this letter, we propose a fusion of wireless and sensor data to enhance beam alignment for cellular-connected UAV massive MIMO communications.
We develop a predictive beamforming framework, including the frame structure and predictive beamformer.
Moreover, we employ an extended Kalman filter (EKF) to integrate channel and sensor data and provide the corresponding state-space and observation models.
Simulation results demonstrate that the proposed scheme can improve position/orientation estimation accuracy significantly, leading to higher spectral efficiency.
\color{black}
\vspace{-1mm}
\end{abstract}

\begin{IEEEkeywords}
Unmanned aerial vehicle (UAV), Multi-input multi-output (MIMO), Integrated sensing and communication (ISAC), Information fusion, Kalman filtering
\end{IEEEkeywords}

\IEEEpeerreviewmaketitle

\vspace{-2mm}

\section{Introduction}

\color{black}
\color{black}
Recently, non-terrestrial networks (NTN) have attracted significant research interest due to their ability to provide ubiquitous coverage in vast areas without the cost of new base station build-out and deployment.
In this context, the 3rd Generation Partnership Project (3GPP) started a study item on NTN since Release 15 to incorporate NTN into cellular architecture \cite{lin20215g}. 
As a key component of NTN, unmanned aerial vehicles (UAVs) are expected to play an essential role in relay applications between a base station and users to mitigate limited cellular coverage in rural areas \cite{zhang2021challenges,zhang2023large}.
\color{black}

The use of massive multiple-input multiple-output (MIMO) technology for cellular-connected UAVs has received a great deal of interest in the literature, particularly given its inherent ability to mitigate interference and increase the spectral efficiency of the network.
\color{black}
Furthermore, massive MIMO is also a key enabler of millimeter-wave/terahertz communications given that a large number of apertures can fit within a small-sized UAV.
Employing a massive MIMO array at the UAV is challenging, given narrow beams and UAV dynamic motion, including rotation. 
The attitude of UAVs constantly changes due to wind gusts and maneuvering.
This leads to significant variations in the angle of arrival (AoA) and departure (AoD) ultimately affecting beamforming solutions derived from stale channel state measures.
\color{blue}

\color{black}

To overcome this problem, it is critical to understand the motion of UAVs.
Previous works have attempted to predict the UAV's motion using onboard sensors such as Global Positioning System (GPS) receivers and inertial measurement units (IMU) for improving beam alignment \cite{zhao2018channel,zhao2018beam}.
However, low-cost GPS/IMU devices on commercial UAVs are prone to measurement errors due to blockage and biases.
Moreover, given potential imperfect mappings between UAV motion and channel, relying solely on onboard sensors may not offer reliable communication.

As an alternative, some studies have sought the assistance of cellular networks for motion prediction  \cite{chang2022integrated,yang2019beam,huang20203d}.
In \cite{chang2022integrated}, radar sensing was employed at the base station to track the UAV position for beam alignment.
In \cite{yang2019beam}, the angular velocity was estimated via pilot transmission at the cellular base station for beam training.
In \cite{huang20203d}, pilot-based AoD prediction and dynamic pilot transmission were investigated to reduce pilot overhead.
However, these works did not take into account the rotation or attitude dynamics of UAVs.
Moreover, radar and pilot transmission may still cause overhead, particularly when the mobility of the UAV is high.

In summary, prior works have focused on predictive beamforming via either onboard sensors \cite{zhao2018channel,zhao2018beam} or network assistance \cite{chang2022integrated,yang2019beam,huang20203d} for UAV beam alignment.
There are limited works on combining the two approaches.
In \cite{wang2021jittering}, GPS/IMU data was used to specify the AoA search range for lower pilot overhead.
In \cite{zhao2023integrated}, the authors proposed an integrated sensing and communication-based channel estimation technique where radar and onboard sensors measure the range and orientation, respectively.
These works focused on using cellular-based and sensor-based information individually.
Nonetheless, the work in \cite{liSmartphoneLocalizationAlgorithm2013} showed that fusing channel and IMU measures can enhance localization precision significantly in indoor scenarios.

Motivated by this, in this paper, we introduce a novel fusion of wireless and sensor data to enhance predictive beamforming in UAV-assisted massive MIMO communications.
The proposed method aims to improve the reliability and precision of beam alignment by complementing GPS/IMU measurements with channel information.
We develop a predictive beamforming framework, including the frame structure and predictive beamformer.
We employ an extended Kalman filter (EKF) to integrate channel and motion data and provide the associated state-space and observation models. 
Simulation results show that the proposed fusion can improve motion tracking significantly, enhancing spectral efficiency.
\color{black}

\color{black}

\vspace{-1mm}

\section{System Model}
\vspace{-2mm}
\subsection{System Setup}
Consider a point-to-point massive MIMO communication system where a base station (BS) serves a UAV.
The BS is equipped with a uniform planar array (UPA) of $N_B=N_{B,h}\times N_{B,v}$ antennas.
The UAV is equipped with a UPA of $N_U=N_{U,h}\times N_{U,v}$ antennas.
The time-varying position, velocity, and attitude of the UAV at time $t$ are denoted by $\textbf{p}(t)=\left[{x}(t),{y}(t),{z}(t)\right]^T$, $\bm{\nu}(t)=\left[\dot{x}(t),\dot{y}(t),\dot{z}(t)\right]^T$, and
\setlength{\skip\footins}{3pt}\footnote{
 We adopt a unit quaternion representation for the UAV attitude since quaternions are free from the well-known Gimbal-lock problem \cite{sidi1997spacecraft}.}$\textbf{q}(t)=\left[q_{1}(t),q_{2}(t),q_{3}(t),q_{4}(t)\right]^T$, respectively.
The BS is stationary at a known position $\textbf{p}_B=\left[x_B,y_B,z_B\right]^T$.

\color{black}

\color{black}
This paper addresses the beam alignment problem to maximize the downlink spectral efficiency.
Thus, we focus on downlink transmission where the BS transmits pilot/data signals to the UAV and the UAV acquires channel state information (CSI) via received pilots.
\color{black}
CSI acquisition at the UAV is done by tracking and predicting the UAV's motion parameters, defined as the position, velocity, acceleration, attitude, and angular rates.
In the proposed framework, the UAV exploits both GPS/IMU measurements and pilots for CSI acquisition.
The UAV then feeds back information about the acquired CSI to the BS to enable beamforming for data transmission.

\begin{figure}[!t]
\center{\includegraphics[width=.900\linewidth]{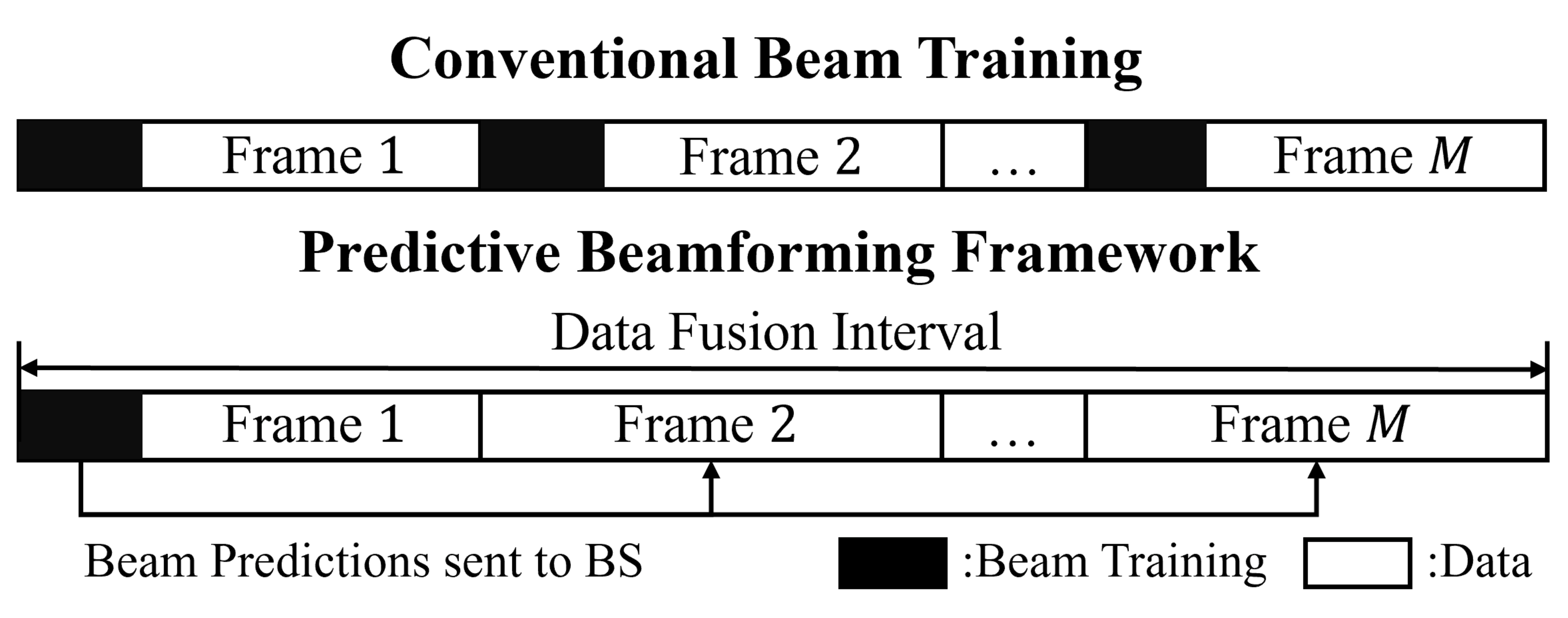}}
\caption{ Proposed downlink frame structure.}\label{fig:Frame}
\vspace{-7mm}
\end{figure}

Our proposed framework employs channel predictions to accomplish the beam alignment task with minimal pilot overhead.
To this end, we consider the frame structure\footnote{{
Conventional beam training requires periodic pilot transmission and feedback, which may cause overhead and latency \cite{love2008overview}.
The considered predictive beamforming can mitigate this problem using UAV motion predictions.
}} illustrated in Fig. \ref{fig:Frame}, where channel predictions are generated at every data fusion interval (DFI) of duration $T_{DFI}$.
Each DFI is composed of $M$ frames, each of which contains $N_s$ symbols of duration $T_{s}$.
The duration of a DFI can be expressed as $T_{DFI}=MT_f$ where $T_f$ is the frame duration with $T_f=N_sT_s$.
During the first frame of each DFI, the BS transmits a burst of $N_p$ pilots to the UAV. 
Once the UAV receives pilots, then it produces AoA/AoD predictions for the subsequent $M-1$ frames.

Following the common assumption in the literature \cite{liu2020radar}, we assume the motion parameters of the UAV remain constant within a frame (e.g., \SI{1}{\milli\second}) but vary from frame to frame.
Accordingly, the position, velocity, and attitude of the UAV at frame $k$ can be expressed as $\textbf{p}_k$, $\bm{\nu}_k$, and $\textbf{q}_k$, respectively.
\color{black}
{Additionally, we assume the UAV's clock is perfectly synchronized to the system clock at the BS\footnote{{
Although we rely on the common perfect synchronization assumption \cite{abu2018error} to gain general insights, the impact of synchronization should be considered.
In practice, synchronization can be done using a two-way protocol or simultaneous localization and synchronization.
}
}.}
\color{black}

\subsection{Pilot Signal Model}
We consider downlink-based channel estimation where the BS transmits pilot signals to the UAV.
The continuous-time $i$th received pilot symbol at the UAV at frame $k$ is written as
\begin{equation}\label{eq:recv_pilot}
    \textbf{y}_{k,i}(t)=\sqrt{P_T}\textbf{H}_k\textbf{f}^p_{i}{s}_{k,i}(t-\tau_k)+\textbf{n}_{k,i}(t),
\end{equation}
where $P_T$ is the BS transmit power, $\textbf{H}_k \in \mathbb{C}^{N_U \times N_B}$ is the BS-to-UAV channel, $\textbf{f}^p_{i}\in{\mathbb{C}^{N_B}}$ is the pilot beamformer with $\lVert \textbf{f}^p_i \rVert_2^2=1$, ${s}_{k,i}(t)\in \mathbb{C}$ is the $i$th pilot symbol, $\tau_k$ is the time-delay, and $\textbf{n}_{k,i}(t) \in \mathbb{C}^{N_U}$ is the Gaussian noise with $\textbf{n}_{k,i}(t) \sim \mathcal{N}(\textbf{0},\sigma^2 \textbf{I})$.
\color{blue}
\color{black}
The transmit pilot signal is given by $s_{k,i}(t)=a_ip(t-iT_{s}-kT_{f})$ where $a_i$ is the $i$th pilot symbol and $p(t)$ is the unit-energy pulse.

\subsection{Channel Model}
Following \cite{zhao2018channel,zhao2018beam,wang2021jittering}, we assume a line-of-sight (LoS) channel for the BS-to-UAV channel, which is given by \cite{abu2018error} \vspace{-1mm}
\color{black}
\begin{equation}
        \textbf{H}_k={\sqrt{N_UN_B}}\alpha_k\textbf{v}_U(\Theta_{U,k},\Phi_{U,k})\textbf{v}_B^H(\Theta_{B,k},\Phi_{B,k}),
\end{equation}
where $\alpha_k$ is the path gain, $\textbf{v}_U(\cdot),\textbf{v}_B(\cdot)$ are the steering vectors of the BS and UAV, respectively, 
$\Theta_{U,k},\Phi_{U,k}$ are the direction cosines of the BS-to-UAV LoS path with respect to the vertical and horizontal axes of the UPA at the UAV, respectively, and $\Theta_{B,k},\Phi_{B,k}$ are the direction cosines of the BS-to-UAV LoS path with respect to the vertical and horizontal axes of the UPA at the BS.
\color{black}
The path gain is given by $\alpha_k=\sqrt{\beta_0}/d_k$ where $\beta_0$ is the path loss at a reference distance (e.g., \SI{1}{\meter}) and $d_k$ is the distance between the BS and UAV with $d_k=\Vert \textbf{p}_k-\textbf{p}_B \Vert$.
The steering vectors for the UPAs of the UAV and the BS are, respectively, given by \cite{wang2021jittering} \vspace{-1mm}
\begin{equation}\label{eq:array_response}
    \begin{aligned}
    \textbf{v}_U(\Theta_{U,k},\Phi_{U,k})=&  \frac{1}{\sqrt{N_U}}  \textbf{b}(\Theta_{U,k},N_{U,v})\otimes\textbf{b}(\Phi_{U,k},N_{U,h}), \\ \nonumber 
    \textbf{v}_B(\Theta_{B,k},\Phi_{B,k})=&  \frac{1}{\sqrt{N_B}} \textbf{b} (\Theta_{B,k},N_{B,v})\otimes \textbf{b}(\Phi_{B,k},N_{B,h}),  \nonumber 
    \end{aligned}
\end{equation}
where $\otimes$ is the Kronecker product and $\textbf{b}(\Theta,N)=e^{\frac{-j\pi(N-1)\Theta}{2}}\big[1,e^{j\pi \Theta}, ..., e^{j\pi(N-1)\Theta}\big]^T$.

\section{UAV Motion Tracking via Channel and GPS/IMU Data Fusion}

\subsection{Antenna Rotation Model}
\begin{figure}[!t]
\label{fig:Rotation}
  \begin{subfigure}[t]{0.18\textwidth}
\center{\includegraphics[width=.98\linewidth]{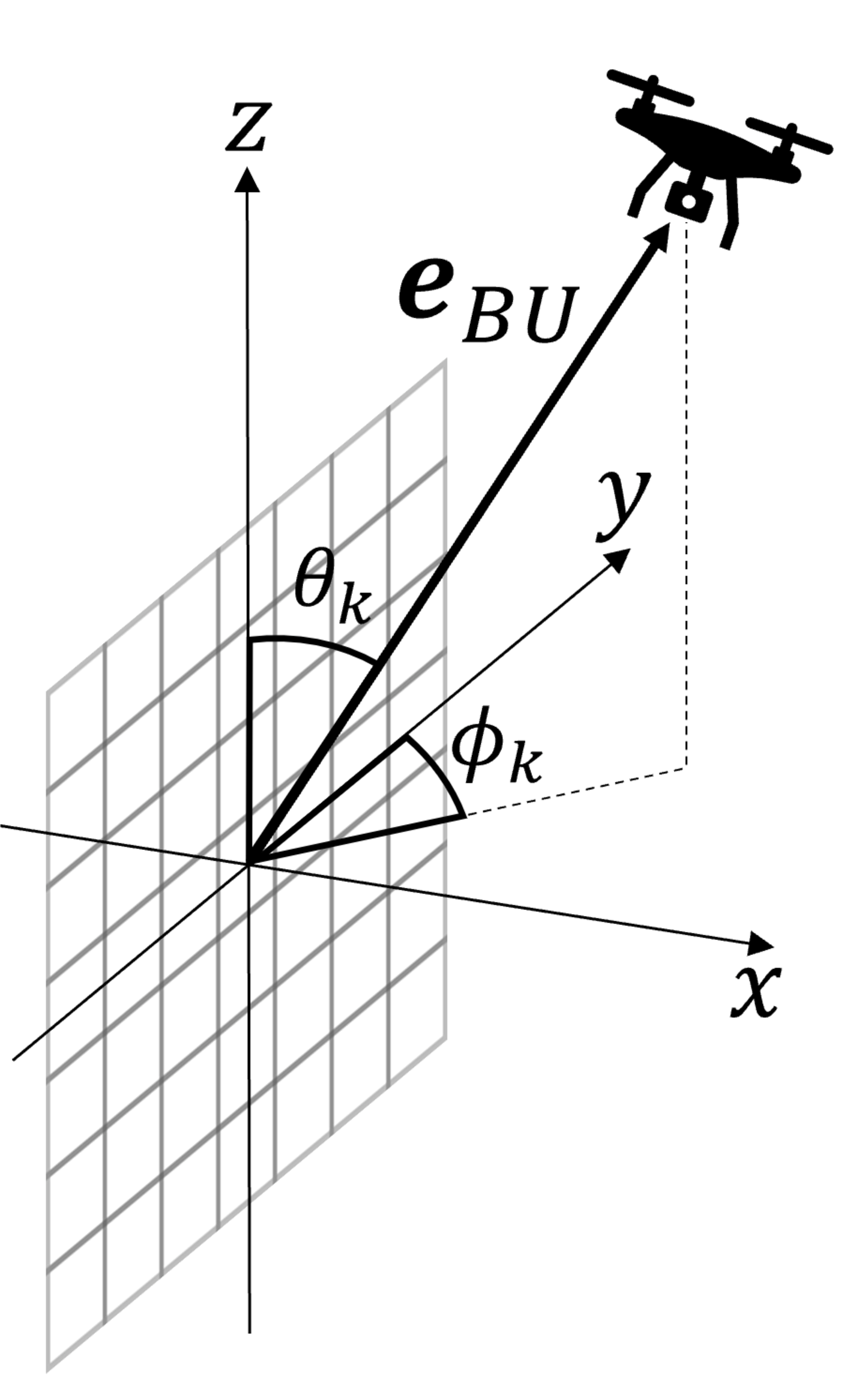}}
\caption{ BS side.}\label{fig:BSgeometry}
 \end{subfigure}
   \begin{subfigure}[t]{0.3\textwidth}
{\includegraphics[width=.995\linewidth]{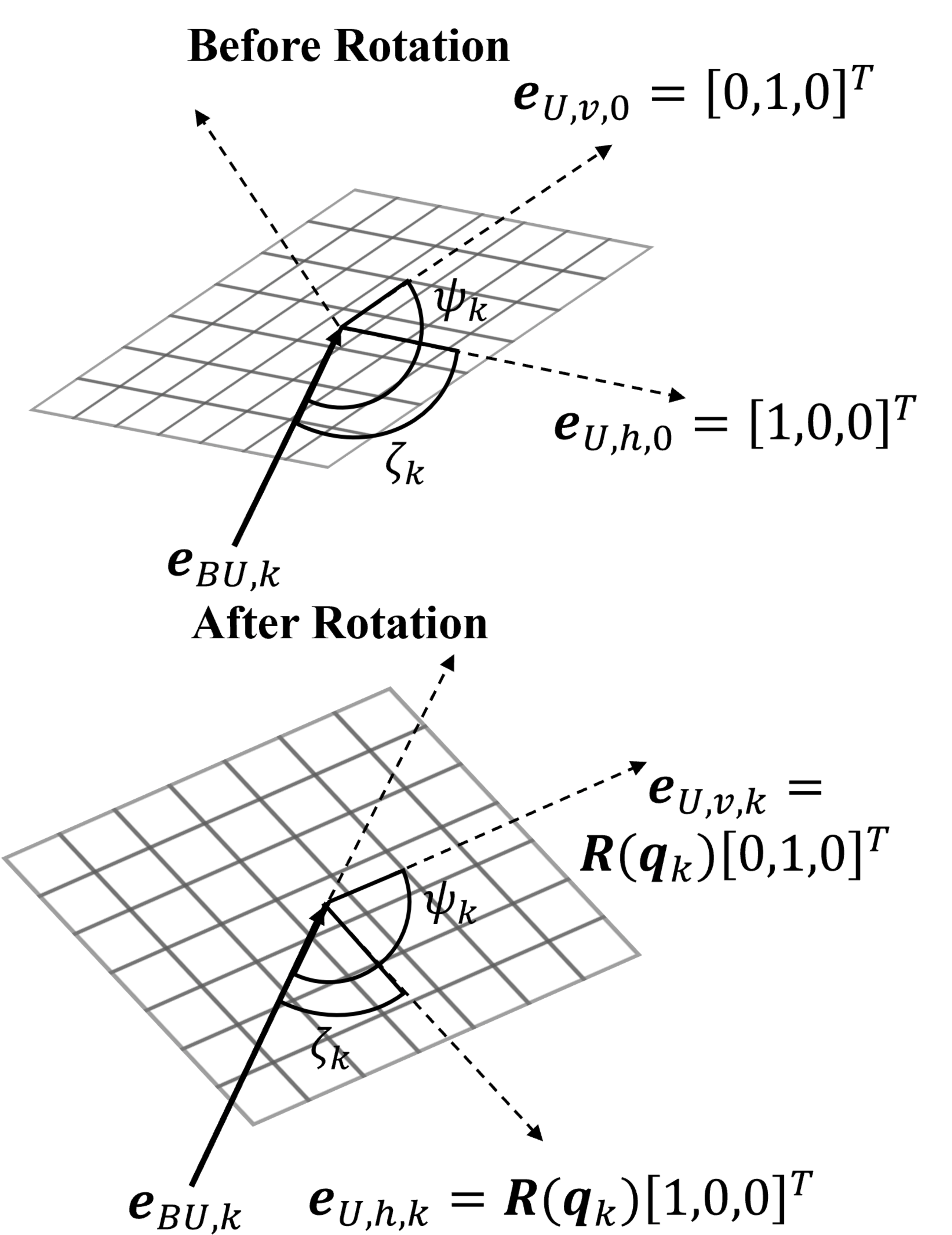}}
\caption{ UAV side.}\label{fig:UAVgeometry}
 \end{subfigure}
 \caption{Channel geometry.}
 \vspace{-6mm}
\end{figure}

Without loss of generality, we assume that the UPA of the BS is positioned at the origin and its horizontal and vertical axes lie in the span of the $y$ and $z$ axes in the Cartesian coordinate system, respectively, as shown in Fig. \ref{fig:BSgeometry}.
\color{black}
Given the BS position and orientation, the unit vector corresponding to the BS-to-UAV LoS path direction is given by \cite{wang2021jittering} \vspace{-1mm}
\begin{equation}
    \textbf{e}_{BU,k}=\textbf{p}_k-\textbf{p}_B=[\sin \phi_k\sin \theta_k, \cos \phi_k\sin \theta_k, \cos \theta_k]^T,
\end{equation}
where $\phi_k$ and $\theta_k$ are the azimuth and elevation angles, respectively, in the Cartesian coordinate system.
The direction cosines of the LoS path with respect to the horizontal and vertical axes of the UPA at the BS are, respectively, given by \vspace{-1mm}
\begin{equation}\label{eq:AOD_BS}
    \begin{aligned}
        \Theta_{B,k}&= [0,0,1]\textbf{e}_{BU,k} =\frac{z_k-z_0}{d_k}, \\  
        \Phi_{B,k}&=  [0,1,0]\textbf{e}_{BU,k} = \frac{y_k-y_0}{d_k}.
    \end{aligned}
\end{equation}
Let $\textbf{e}_{U,h,k}$ and $\textbf{e}_{U,v,k}$ be the vectors corresponding to the horizontal and vertical axes of the UPA of the UAV, respectively.
Without loss of generality, we set the initial horizontal and vertical axes vectors of the UPA at the UAV  without rotation as $\textbf{e}_{U,h,0}=[1,0,0]^T$ and $\textbf{e}_{U,v,0}=[0,1,0]^T$, respectively.
\color{black}
In practice, the horizontal and vertical axes $\textbf{e}_{U,h,k}$, $\textbf{e}_{U,v,k}$ of the UPA at the UAV, respectively, rotate dynamically according to the attitude $\textbf{q}_k$ of the UAV.
Given an attitude vector $\textbf{q}=[q_1,q_2,q_3,q_4]$, the rotation matrix for the transformation from the body frame to the navigation frame is given by \cite{sidi1997spacecraft}
\begin{equation*}
    \begin{aligned}
        &\textbf{R}(\textbf{q})=\\ 
    &\setlength\arraycolsep{0.1pt}
    \begin{bmatrix}
    q_{4}^2+q_{1}^2-q_{2}^2-q_{3}^2 & 2(q_{1}q_{2}-q_{3}q_{4}) &2(q_{1}q_{3}+q_{2}q_{4}) \\[0.01pt]
    2(q_{1}q_{2}+q_{3}q_{4}) & q_{4}^2-q_{1}^2+q_{2}^2-q_{3}^2 & 2(q_{2}q_{3}-q_{1}q_{4}) \\[0.01pt]
    2(q_{1}q_{3}-q_{2}q_{4}) & 2(q_{2}q_{3}+q_{1}q_{4}) & q_{4}^2-q_{1}^2-q_{2}^2+q_{3}^2
    \end{bmatrix}.
    \end{aligned}
\end{equation*}
We use $\zeta_k$ and $\psi_k$ to denote the AoAs at the UAV with respect to the horizontal and vertical axes of the UPA at the UAV, respectively, as illustrated in Fig. \ref{fig:UAVgeometry}.
The direction cosines of $\textbf{e}_{BU,k}$ with respect to the horizontal and vertical axes of the UPA at the UAV are, respectively, given by 
\cite{wang2021jittering} \vspace{-1.5mm}
\begin{equation}\label{eq:AOA_UAV}
    \begin{aligned}
        \Phi_{U,k}&=\cos{\zeta_k}=
       \textbf{e}^T_{BU,k}\textbf{e}_{U,h,k}=\textbf{e}^T_{BU,k}\textbf{R}(\textbf{q}_k)\textbf{e}_{U,h,0}, \\ 
        \Theta_{U,k}&=\cos{\psi_k}=
       \textbf{e}^T_{BU,k}\textbf{e}_{U,v,k}=\textbf{e}^T_{BU,k}\textbf{R}(\textbf{q}_k)\textbf{e}_{U,v,0}.
    \end{aligned}
\end{equation}
\vspace{-3mm}

\subsection{State-space Model}

The UAV motion state vector is given by $\textbf{x}_k=
\left[\textbf{p}^T_k,{\bm{\nu}}^T_k,{\textbf{a}}^T_k,\textbf{q}^T_k,\bm{\omega}^T_k\right]^T$
where ${\textbf{a}}_k=[\ddot{x}_k,\ddot{y}_k,\ddot{z}_k]^T$ is the acceleration vector in the \textit{x/y/z} axes and $\bm{\omega}_k=[\omega_{1,k},\omega_{2,k},\omega_{3,k}]^T$ is the angular rates with respect to the UAV body frame. 
The state transition model is given by \cite{bar2001estimation,sidi1997spacecraft}  \vspace{-1mm}
\begin{equation}\label{eq:transition}
    \begin{aligned}
        \textbf{p}_k&=\textbf{p}_{k-1}+\bm{\nu}_{k-1}T_f+\frac{1}{2}\textbf{a}_{k-1}T_f^2, \\[0.01pt] 
\textbf{q}_k&=\left(\textbf{I}_4+\frac{1}{2}T_f\bm{\Omega}(\bm{\omega}_{k-1})\right)\textbf{q}_{k-1}, \\[0.01pt]  
\bm{\nu}_k&=\bm{\nu}_{k-1}+\textbf{a}_{k-1}T_f, \ 
\textbf{a}_{k}=\textbf{a}_{k-1}, \
 \bm{\omega}_k=\bm{\omega}_{k-1},
    \end{aligned}
\end{equation}
with 
\begin{equation}
    \begin{aligned}
            \bm{\Omega}(\bm{\omega}_k)=
            \setlength\arraycolsep{1pt}\begin{bmatrix}
        0 &\omega_{3,k}&-\omega_{2,k} & \omega_{1,k} \\[0.01pt]
        -\omega_{3,k}&0 &\omega_{1,k} & \omega_{2,k} \\[0.01pt]
        \omega_{2,k}&-\omega_{1,k}&0&\omega_{3,k} \\[0.01pt]
        -\omega_{1,k}&-\omega_{2,k}&-\omega_{3,k}&0
        \end{bmatrix}.
    \end{aligned}
\end{equation}
From the transition model, the state-space model for the UAV motion state can be expressed as \vspace{-1mm}
\begin{equation}\label{eq:dynamic}
        \textbf{x}_k=\bm{\eta}(\textbf{x}_{k-1}) + \textbf{u}_k,
\end{equation}
where $\bm{\eta}(\cdot)$ is the non-linear state transition model defined in \eqref{eq:transition} and $\textbf{u}_k$ is the process noise vector with $\textbf{u}_k\sim \mathcal{N}(\textbf{0},\,\textbf{U}_k)$.
The process noise covariance is given by $\textbf{U}_{k}=\mathrm{blkdiag}\left( \sigma_{1}^2\bm{\Gamma}\otimes\textbf{I}_3,\sigma_{2}^2T_f\bm{\Xi}_k\bm{\Xi}^T_k, \sigma_{2}^2T_f\textbf{I}_3 \right)$
where $\sigma_{1}^2$ and $\sigma_{2}^2$ are the jerk and angular acceleration noise intensities, respectively.
The matrices $\bm{\Gamma},\bm{\Xi}_k$ are, respectively, given by \cite{bar2001estimation,lefferts1982kalman}\vspace{-2mm}
\begin{align*}
\bm{\Gamma}&=
\begin{bmatrix}
        \frac{1}{20}T_f^5 & \frac{1}{8}T_f^4 & \frac{1}{6}T_f^3 \\
        \frac{1}{8}T_f^4 & \frac{1}{3}T_f^3 & \frac{1}{2}T_f^2 \\
        \frac{1}{6}T_f^3 & \frac{1}{2}T_f^2 & T_f \\
    \end{bmatrix}, \bm{\Xi}_k=    
    \frac{T_f}{2}
    \setlength\arraycolsep{0.5pt}\begin{bmatrix}
        q_{4,k}&-q_{3,k}&q_{2,k} \\[0.001pt]
        q_{3,k}&q_{4,k}&-q_{1,k} \\[0.001pt]
        -q_{2,k}&q_{1,k}&q_{4,k} \\[0.001pt]
        -q_{1,k}&-q_{2,k}&-q_{3,k} 
    \end{bmatrix}.
\end{align*}
\vspace{-4mm}
\subsection{Observation Model}
At the first frame of each DFI the UAV acquires GPS/IMU and channel parameter measurements. 
Note that the index of the first frame of the $\ell$th DFI is $\ell M$.
The nonlinear observation function of the proposed data fusion is obtained as \vspace{-2mm}
\begin{equation}\label{eq:obrv_model}
    \hat{\textbf{r}}_{\ell M}=[\hat{\textbf{r}}^T_{gps,\ell M},\hat{\textbf{r}}^T_{imu,\ell M},\hat{\textbf{r}}^T_{ch,\ell M}]^T=\bm{g}(\textbf{x}_{\ell M})+\textbf{z}_{\ell M},
\end{equation}
where $\hat{\textbf{r}}_{gps,\ell M}=[\hat{{\textbf{p}}}^T_{\ell M},\hat{\bm{\nu}}^T_{\ell M}]^T$, $\hat{\textbf{r}}_{imu,\ell M}=[\hat{\textbf{a}}^T_{b,\ell M},\hat{\bm{\omega}}^T_{\ell M}]^T$, $\hat{\textbf{r}}_{ch,\ell M}$ are the GPS, IMU, and channel observation vectors, respectively, and $\textbf{z}_{\ell M}$ is the observation noise.
The GPS/IMU observation model is given by \cite{liSmartphoneLocalizationAlgorithm2013,prieto2016context}\vspace{-2mm}
\begin{equation}\label{eq:GPS_IMU_obrv}
    \begin{aligned}
        \hat{\textbf{p}}_{\ell M} 
    &=\textbf{p}_{\ell M}+\textbf{z}_{p,\ell M},
    \hat{\bm{\nu}}_{\ell M} 
    =\bm{\nu}_{\ell M}+\textbf{z}_{\nu,\ell M}, \\
    \hat{\textbf{a}}_{b,\ell M} &  =\textbf{R}^T(\textbf{q}_{\ell M})(\textbf{a}_{\ell M}-\textbf{a}_g)+\textbf{z}_{a,\ell M},  \\ 
    \hat{\bm{\omega}}_{\ell M} &  =\bm{\omega}_{\ell M}+\textbf{z}_{\omega,\ell M},
    \end{aligned} \vspace{-2mm}
\end{equation}
where $\hat{\textbf{p}}_{\ell M},\hat{\bm{\nu}}_{\ell M}\in\mathbb{R}^3$ are the GPS position and velocity measurements with respect to the navigation frame, respectively, $\hat{\textbf{a}}_{b,\ell M},\hat{\bm{\omega}}_{\ell M}\in\mathbb{R}^3$ are the IMU acceleration and angular rate measurements with respect to the UAV body frame, respectively, $\textbf{a}_g=[0,0,9.81]^T$\si{m/s^{2}} is the gravity acceleration, and $\textbf{z}_{nav,\ell M}=[{\textbf{z}}^T_{p,\ell M},{\textbf{z}}^T_{\nu,\ell M},{\textbf{z}}^T_{a,\ell M},{\textbf{z}}^T_{\omega,\ell M}]^T$ is the GPS/IMU observation noise with $\textbf{z}_{nav,\ell M}\sim\mathcal{N}(\textbf{0},\text{diag}(\sigma^2_p,\sigma^2_{\nu},\sigma^2_a,\sigma^2_{\omega})\otimes \textbf{I}_3$).
\color{black}
At each DFI, the UAV estimates the channel parameters\footnote{{In practice, channel parameters can be estimated using Kalman filtering or compressive sensing \cite{wang2021jittering}, which is beyond the scope of this paper.
}} from the received pilots in \eqref{eq:recv_pilot}.
The channel parameter observation function is given by \vspace{-2mm}
\begin{equation}\label{eq:channel_obrv}
    \begin{aligned}
        \hat{\textbf{r}}_{ch,\ell M}&=\bm{g}_{ch}(\textbf{x}_{\ell M})+\textbf{z}_{ch,\ell M} \\ 
    &=[\Theta_{B,\ell M},\Phi_{B,\ell M},\Theta_{U,\ell M},\Phi_{U,\ell M},\tau_{\ell M}]+\textbf{z}_{ch,\ell M},
    \end{aligned}
\end{equation}
where 
$\textbf{z}_{ch,\ell M}$ is the observation noise for the channel parameters with $\textbf{z}_{ch,\ell M} \sim \mathcal{N}(\textbf{0},\textbf{V}_{ch,\ell M})$.
The observation noise covariance for the channel parameters can be approximated\footnote{\color{black}We assume a high SNR condition owing to the LoS channel of the UAV. Under these circumstances, maximum likelihood estimation is asymptotically efficient, and thus the mean square error (MSE) approaches the CRB \cite{kay1993fundamentals}.}
with the Cramer-Rao lower bound (CRB) for the channel parameters, which can be obtained as  $\textbf{V}_{ch,\ell M}=\textbf{J}^{-1}_{\ell M}$ 
where $\textbf{J}_{\ell M}\in\mathbb{R}^{5\times5}$ is the Fisher information matrix (FIM) for the channel parameters \cite{kay1993fundamentals,abu2018error} (See Appendix 
 \ref{sec:appendix_B} for details).

\begin{figure*}[!t]
\label{fig:PE_AE}
  \begin{subfigure}[t]{0.33\textwidth}  
    \centering 
  \includegraphics[width=0.99\linewidth]{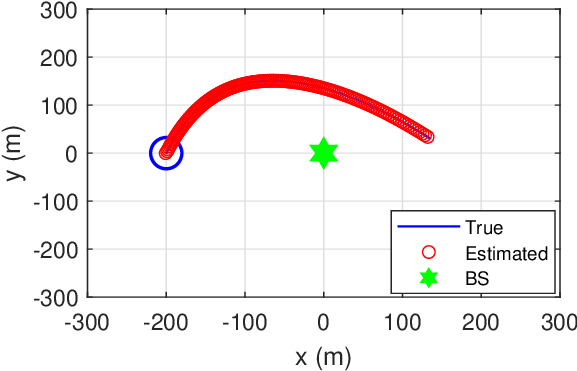}
  \caption{UAV trajectory (top view)}
     \label{fig:TJ}
     \vspace{-1mm}
 \end{subfigure}
   \begin{subfigure}[t]{0.33\textwidth}
     \centering
     \includegraphics[width=0.99\linewidth]{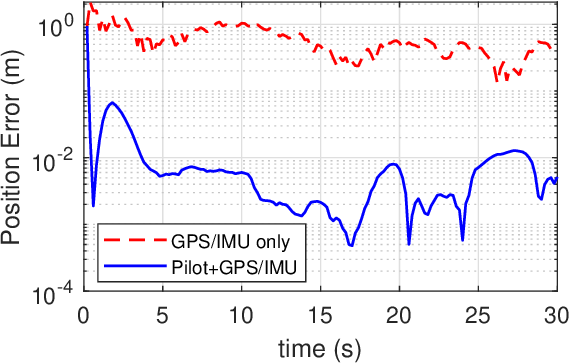}
     \caption{Position error (\si{\meter})}
     \label{fig:PE}
     \vspace{-1mm}
 \end{subfigure}
      \begin{subfigure}[t]{0.33\textwidth}
         \centering
         \includegraphics[width=0.99\linewidth]{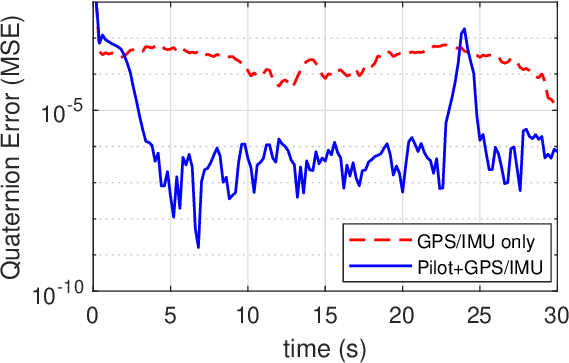}
         \caption{Attitude error (MSE)}
         \label{fig:AE}
         \vspace{-1mm}
     \end{subfigure}
     \caption{UAV trajectory, position error, and attitude error in the simulation}
     \vspace{-6.5mm}
\end{figure*}

  

\subsection{Data Fusion with EKF}
To address the non-linearity in the state-space model \eqref{eq:dynamic} and observation model \eqref{eq:obrv_model}, we adopt an EKF\footnote{Although this paper focuses on an EKF, any type of non-linear filter such as an unscented Kalman filter and particle filter can be applied to our method. 
A practical realization of this method can be a bank of non-linear filters such as interacting multiple model (IMM) filters to improve accuracy and robustness.
} for predicting and updating the state and covariance matrices.
The state and covariance are updated at every DFI with intervals of $M$ frames.
The $m$-step state and covariance predictions for $m=1,\dots,M$ at the $\ell$th DFI are, respectively, given by \vspace{-1.5mm}
\begin{equation}\label{eq:predict_state_cov}
    \begin{aligned}
    \hat{\textbf{x}}_{\ell M+m|\ell M} &= \bm{\eta}(\hat{\textbf{x}}_{\ell M+m-1|\ell M}), \\ 
    \textbf{P}_{\ell M+m|\ell M} &=\textbf{F}_{\ell M+m}\textbf{P}_{\ell M+m-1|\ell M}\textbf{F}^T_{\ell M+m}+\textbf{U}_{\ell M+m-1},
\end{aligned}
\end{equation}
where $\textbf{F}_{\ell M +m}=\frac{\partial \bm{\eta}}{\partial \textbf{x}}\big|_{\textbf{x}=\hat{\textbf{x}}_{\ell M +m-1|\ell M}}$ is the Jacobian matrix of the state-space model,
and $\hat{\textbf{x}}_{\ell M+m-1|\ell M}$, $\textbf{P}_{\ell M+m-1|\ell M}$ are the $(m-1)$-step state and covariance predictions. 
The updated state and covariance at the $\ell$th DFI are, respectively, given by \vspace{-2mm}
\begin{equation}\label{eq:mean_cov_est}
    \begin{aligned}
    \hat{\textbf{x}}_{(\ell+1) M|(\ell+1) M} &= \hat{\textbf{x}}_{(\ell+1) M|\ell M} \\ 
    &+ \textbf{K}_{(\ell+1) M}\left(\hat{\textbf{r}}_{(\ell+1) M}-\bm{g}(\hat{\textbf{x}}_{(\ell+1) M|\ell M})\right), \\
    \textbf{P}_{(\ell+1) M|(\ell+1) M} &= \big(\textbf{I}-\textbf{K}_{(\ell+1) M}\textbf{G}_{(\ell+1) M}\big)  \textbf{P}_{(\ell+1) M|\ell M},
\end{aligned}
\end{equation}
where $\textbf{K}_{(\ell+1) M}$ is the Kalman gain, which is given by \vspace{-2mm}
\begin{equation}
    \begin{aligned}
            \textbf{K}_{(\ell+1) M} &= \textbf{P}_{(\ell+1) M|\ell M}\textbf{G}^T_{(\ell+1) M} 
        \\ \nonumber
        &\cdot\big(\textbf{V}_{(\ell+1) M} +  \textbf{G}_{(\ell+1) M}  \textbf{P}_{(\ell+1) M|\ell M}\textbf{G}^T_{(\ell+1) M}\big)^{-1},
        \vspace{-1.5mm}
    \end{aligned}
\end{equation}
where $\textbf{G}_{(\ell+1) M}$ is the Jacobian of the observation model with $\textbf{G}_{(\ell+1) M} = \frac{\partial \bm{g}}{\partial \textbf{x}} \big\vert_{\textbf{x}=\hat{\textbf{x}}_{(\ell+1) M|\ell M}}$.

\textbf{EKF Complexity Analysis:} Let $L_x$ and $L_r$ be the dimensions of the state and observation vectors, respectively.
The $M$-step state and covariance predictions cost $O(L_x^2M)$ and $O(L_x^3M)$, respectively.
The Kalman gain calculation costs $O(L_x^2L_r+L_xL_r^2+L_r^3)$.
The state and covariance updates cost $O(L_xL_r)$ and $O(L_xL_r^2+L_x^3)$, respectively.
\color{black}

\section{Predictive Beamforming}
\subsection{Predictive Beamforming and Combining}

Based on the obtained motion parameters \eqref{eq:mean_cov_est}, the UAV predicts the AoA/AoD to formulate the beamformer/combiner for data transmission.
According to the observation model in \eqref{eq:channel_obrv} and the state prediction in \eqref{eq:predict_state_cov}, the $m$-step AoA/AoD prediction at the $\ell$th DFI is given by \vspace{-1.5mm}
\begin{align*}
    &[\bm{g}_{ch}(\hat{\textbf{x}}_{\ell M+m|\ell M})]_{1:4} \\ \nonumber
    &=    [\hat{\Theta}_{B,\ell M+m|\ell M},\hat{\Phi}_{B,\ell M+m|\ell M},\hat{\Theta}_{U,\ell M+m|\ell M},\hat{\Phi}_{U,\ell M+m|\ell M}], 
\end{align*}
where $[\cdot]_{1:4}$ returns the first-to-fourth entries of a vector, 
$\hat{\Theta}_{B,\ell M+m|\ell M},\hat{\Phi}_{B,\ell M+m|\ell M}$ are the $m$-step predictions for the BS-side direction cosines, and
$\hat{\Theta}_{U,\ell M+m|\ell M},\hat{\Phi}_{U,\ell M+m|\ell M}$ are the $m$-step predictions\footnote{For the BS to determine the data beamformer, the predictions should be fed back to the BS. 
Although quantized feedback is typically used in practical systems \cite{love2008overview}, this paper assumes complete feedback is available for simplicity.}
 for the UAV-side direction cosines.

The $m$-step predictive beamformer and combiner can be obtained by plugging the direction cosine predictions into the steering vectors as \vspace{-2mm}
\begin{equation}
    \begin{aligned}
    \textbf{f}_{\ell M+m}&=\textbf{a}_B\left(\hat{\Theta}_{B,\ell M+m|\ell M},\hat{\Phi}_{B,\ell M+m|\ell M}\right), \\
    \textbf{w}_{\ell M+m}&=\textbf{a}_U\left(\hat{\Theta}_{U,\ell M+m|\ell M},\hat{\Phi}_{U,\ell M+m|\ell M}\right).
\end{aligned}
\end{equation}



\subsection{Spectral Efficiency}
In the data transmission phase at frame $k$, the $n$th received data symbol at the UAV is given by
\begin{equation}
    {y}_{k,n}(t)=\sqrt{P_T}\textbf{w}^H_k\textbf{H}_k\textbf{f}_ks^d_{k,n}(t-\tau_k)+\textbf{w}^H_k\textbf{n}_{k,n}(t),
\end{equation}
where 
$s^d_{k,n}(t)$ is the $n$th data symbol, and $\textbf{n}_{k,n}(t)$ is the noise with $\textbf{n}_{k,n}(t) \sim \mathcal{N}(\textbf{0},\sigma^2 \textbf{I})$.
The signal-to-noise ratio (SNR) at the UAV at frame $k$ is given by $\gamma_k={\lambda_k|\textbf{w}^H_k\textbf{v}_U(\Phi_{U,k},\Theta_{U,k})\textbf{v}^H_B(\Phi_{B,k},\Theta_{B,k})\textbf{f}_k|^2}$ where $\lambda_k=P_TN_UN_B\alpha_k^2/\sigma^2$.
The spectral efficiency of the BS-UAV link at frame $k$ is given by $R_k=\log_2(1+\gamma_k) \  \mathrm{bps/\si{\hertz}}$.
\section{Simulation Results}
In this section, we present our simulation results.
In our setup, the BS is located at the origin and the UAV departs from an initial point $\textbf{p}_0=[-200,0,100]\si{m}$ with an initial speed \SI{70}{km/h}.
The UAV flies for 30 seconds and its trajectory is randomly generated according to the state-space and process noise models, as depicted in Fig. \ref{fig:TJ}.
\color{black}
We set $f_c=\SI{30}{\giga\hertz}$, $\beta_0=10^{-6.2}$, $N_P=N_B$, $T_f=\SI{1}{\ms}$, $T_{DFI}=\SI{200}{\ms}$, and $M=200$.
The GPS/IMU measurement noise parameters are set to $\sigma_{p}=\SI{3}{m}$, $\sigma_{v}=\SI{0.03}{m/s}$, $\sigma_{a}=\SI{2e-3}{m/s^2/\sqrt{\hertz}}$, and $\sigma_{\omega}=\SI{5.2e-4}{\radian/\second}$ \cite{liSmartphoneLocalizationAlgorithm2013}.
The process noise parameters are set to $\sigma_1=\SI{2.24e-2}{m/s^3}$ and $\sigma_2=\SI{0.1}{\radian/\second^2}$.
\color{black}
We use the GPS/IMU-only and pilot-only schemes as baselines.
For the GPS/IMU-only scheme, an EKF was applied to track the motion parameters.
The pilot-only case uses the same channel estimates within a DFI.

Fig. \ref{fig:PE} and \ref{fig:AE} plot the position and attitude errors of the proposed fusion and GPS/IMU-only scheme with $16\times16$ UPAs at the BS and UAV and BS power $P_T=10\ dBm$.
The time-averaged position and attitude errors of the proposed scheme are $\SI{0.014}{m}$ and $\SI{1.4e-4}{}$, respectively, whereas those of the GPS/IMU-only baseline are $\SI{0.607}{m}$ and $\SI{3.6e-4}{}$, respectively.
It can be observed the proposed data fusion enhanced the position and attitude tracking performances considerably.
This can be attributed to the inclusion of high-resolution angle estimates, obtained from the massive number of antennas.
\color{black}

\begin{figure}[!t]
\center{\includegraphics[width=.7\linewidth]{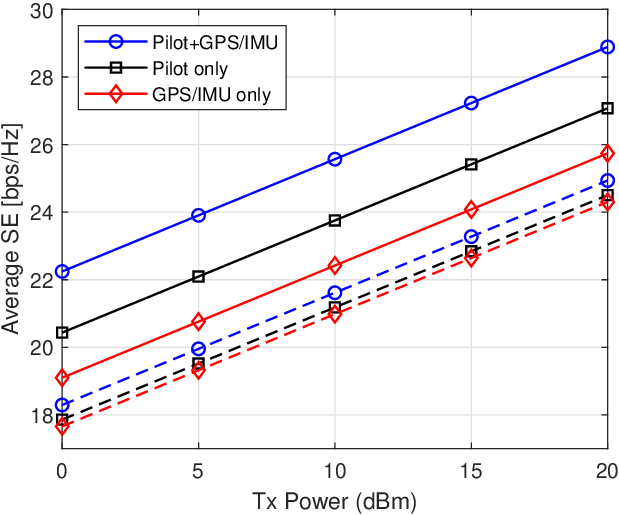}}
\caption{ Average spectral efficiency versus Tx power. Solid lines and dashed lines correspond to $32\times 32$ and $16 \times 16$ UPAs, respectively.}\label{fig:SE_Power}
\vspace{-6mm}
\end{figure}

Fig. \ref{fig:SE_Power} plots the time-averaged spectral efficiencies for two antenna configurations $16\times 16$ and $32\times 32$ UPAs at both the BS and UAV with increasing BS powers from \SI{0}{\dBm} to \SI{20}{\dBm}.
In all cases, the proposed data fusion outperforms the baselines due to the more accurate AoA/AoD predictions than the baselines.
The performance gain is mainly due to the reduced motion prediction error achieved by channel and GPS/IMU information fusion. 
It is shown that the spectral efficiencies are higher with the $32 \times 32$ UPA than those with the $16\times 16$ UPA due to the higher array gain.
In addition, the performance gain of the proposed method is larger with the higher number of antennas.
This implies the impact of channel prediction accuracy is higher when the beam is narrower.
\color{black}

\section{Conclusion}
\vspace{-1mm}
In this paper, we investigated a novel fusion of channel and GPS/IMU data for predictive beamforming in UAV-assisted massive MIMO communications.
We developed an EKF-based data fusion method that can improve significantly the motion tracking and prediction accuracy compared to the GPS/IMU-only case.
Simulation results showed the effectiveness of the proposed scheme, particularly with the massive number of antennas.
As a byproduct of the proposed data fusion, the UAV is allowed to refine the motion parameters, which improves the maneuvering behavior of the UAV as well as communication. 
\color{blue}
\color{black}


\appendices
\section{CRB Derivation}\label{sec:appendix_B}
\vspace{-1.5mm}
The FIM of the channel parameters at frame $\ell M$ is given by $\textbf{J}_{\ell M}=\sum_{i=1}^{N_p}\textbf{J}^e_{\ell M}(i)\Big|_{\bm{g}_{ch}(\textbf{x}_{\ell M})}$ 
 where $\textbf{J}^e_{\ell M}(i)$ is the equivalent FIM (EFIM) of the channel parameters in the $i$th pilot symbol \cite{abu2018error}.
 For brevity, we temporarily drop the frame indices.
The EFIM for the $i$th pilot symbol is given by \cite{abu2018error} \vspace{-1mm}
\begin{equation*}
    \textbf{J}^e(i)=\begin{bmatrix}
        \textbf{J}^e_U(i) & \textbf{0}_{2\times2} & \textbf{0}_{2\times1} \\[0.001pt]
        \textbf{0}_{2\times2} & \textbf{J}^e_B(i) & \textbf{0}_{2\times1}\\[0.001pt]
        \textbf{0}_{1\times 2}&\textbf{0}_{1\times 2} & 8\pi^2\lambda B_{eff}^2|\textbf{v}_B^H(\Theta_B,\Phi_B) \textbf{f}^p_i|^2
    \end{bmatrix},
\end{equation*}
where $\textbf{J}^e_U(i),\textbf{J}^e_B(i)$ are the EFIMs for the cosines of the AoAs at the UAV and the cosines of the AoDs at the BS, respectively, 
$B_{eff}$ is the effective bandwidth with $B_{eff}=\sqrt{\int^{B/2}_{-B/2} f^2|P(f)|^2df}$, $B$ is the bandwidth, and $|P(f)|^2$ is the power spectral density of a unit-energy pulse.
The EFIMs for the cosines of the AoAs and AoDs are, respectively, given by \vspace{-2mm}
\begin{equation*}
    \begin{aligned}
    \textbf{J}^e_U(i)= &
    2G\lambda\begin{bmatrix}
                 \Vert \textbf{k}_U \Vert_2^2   &  \Re\{\textbf{k}^H_U\bm{\varphi}_U\}\\[0.1pt]
         \Re\{\textbf{k}^H_U\bm{\varphi}_U\}  &  \Vert \bm{\varphi}_U \Vert_2^2\\[0.01pt]
    \end{bmatrix}, \\
        \textbf{J}^e_B(i)= &
    2\lambda\setlength\arraycolsep{3pt}\begin{bmatrix}
         {\left|\textbf{k}^H_B \textbf{f}^p_i\right|^2-\frac{1}{G}\xi_{\Theta}^2}  &  \Re\{\chi-\frac{1}{G}\xi_{\Phi}\xi_{\Theta}\}\\
       \Re\{\chi-\frac{1}{G}\xi_{\Phi}\xi_{\Theta}\}& {\left|\bm{\varphi}^H_B \textbf{f}^p_i\right|^2-\frac{1}{G}\xi_{\Phi}^2} 
    \end{bmatrix}, \vspace{-1mm}
\end{aligned}
\end{equation*}
where  \vspace{-1mm}
\begin{equation*}
    \begin{aligned}
    \textbf{k}_U& \triangleq\frac{\partial \textbf{v}_U}{\partial \Theta_U}, \
    \bm{\varphi}_U \triangleq\frac{\partial \textbf{v}_U}{\partial \Phi_U}, 
    \textbf{k}_B \triangleq\frac{\partial \textbf{v}_B}{\partial \Theta_B}, \ \bm{\varphi}_B \triangleq\frac{\partial \textbf{v}_B}{\partial \Phi_B},
    \\ \nonumber
    \xi_{\Theta}&\triangleq\Re\{\textbf{k}^H_B\textbf{f}^p_i(\textbf{f}^p_i)^H\textbf{v}_B\},\ \xi_{\Phi}\triangleq\Re\{\bm{\varphi}^H_B\textbf{f}^p_i(\textbf{f}^p_i)^H\textbf{v}_B\}, \\ \nonumber
    \chi & \triangleq \textbf{k}^H_B\textbf{f}^p_i(\textbf{f}^p_i)^H\bm{\varphi}_B.
\end{aligned}
\end{equation*}

\color{black}

\vspace{-2mm}
\bibliographystyle{IEEEtran}
\bibliography{IEEEabrv,references}

\begin{thebibliography}{10}
\providecommand{\url}[1]{#1}
\csname url@samestyle\endcsname
\providecommand{\newblock}{\relax}
\providecommand{\bibinfo}[2]{#2}
\providecommand{\BIBentrySTDinterwordspacing}{\spaceskip=0pt\relax}
\providecommand{\BIBentryALTinterwordstretchfactor}{4}
\providecommand{\BIBentryALTinterwordspacing}{\spaceskip=\fontdimen2\font plus
\BIBentryALTinterwordstretchfactor\fontdimen3\font minus \fontdimen4\font\relax}
\providecommand{\BIBforeignlanguage}[2]{{%
\expandafter\ifx\csname l@#1\endcsname\relax
\typeout{** WARNING: IEEEtran.bst: No hyphenation pattern has been}%
\typeout{** loaded for the language `#1'. Using the pattern for}%
\typeout{** the default language instead.}%
\else
\language=\csname l@#1\endcsname
\fi
#2}}
\providecommand{\BIBdecl}{\relax}
\BIBdecl

\bibitem{lin20215g}
X.~Lin, S.~Rommer, S.~Euler, E.~A. Yavuz, and R.~S. Karlsson, ``{5G} from space: An overview of {3GPP} non-terrestrial networks,'' \emph{IEEE Commun. Stand. Mag.}, vol.~5, no.~4, pp. 147--153, 2021.

\bibitem{zhang2021challenges}
Y.~Zhang, D.~J. Love, J.~V. Krogmeier, C.~R. Anderson, R.~W. Heath, and D.~R. Buckmaster, ``Challenges and opportunities of future rural wireless communications,'' \emph{{IEEE} Commun. Mag.}, vol.~59, no.~12, pp. 16--22, 2021.

\bibitem{zhang2023large}
Y.~Zhang, J.~V. Krogmeier, C.~R. Anderson, and D.~J. Love, ``Large-scale cellular coverage simulation and analyses for follow-me {UAV} data relay,'' \emph{{IEEE} Trans. Wireless Commun.}, 2023.

\bibitem{zhao2018channel}
J.~Zhao, F.~Gao, L.~Kuang, Q.~Wu, and W.~Jia, ``Channel tracking with flight control system for {UAV} {mmWave MIMO} communications,'' \emph{{IEEE} Commun. Lett.}, vol.~22, no.~6, pp. 1224--1227, 2018.

\bibitem{zhao2018beam}
J.~Zhao, F.~Gao, Q.~Wu, S.~Jin, Y.~Wu, and W.~Jia, ``Beam tracking for {UAV} mounted satcom on-the-move with massive antenna array,'' \emph{{IEEE} J. Sel. Areas Commun.}, vol.~36, no.~2, pp. 363--375, 2018.

\bibitem{chang2022integrated}
B.~Chang, W.~Tang, X.~Yan, X.~Tong, and Z.~Chen, ``Integrated scheduling of sensing, communication, and control for mm{W}ave/{THz} communications in cellular connected {UAV} networks,'' \emph{{IEEE} J. Sel. Areas Commun.}, 2022.

\bibitem{yang2019beam}
L.~Yang and W.~Zhang, ``Beam tracking and optimization for uav communications,'' \emph{{IEEE} Trans. Wireless Commun.}, vol.~18, no.~11, pp. 5367--5379, 2019.

\bibitem{huang20203d}
Y.~Huang, Q.~Wu, T.~Wang, G.~Zhou, and R.~Zhang, ``{3D} beam tracking for cellular-connected {UAV},'' \emph{{IEEE} Wireless Commun. Lett.}, vol.~9, no.~5, pp. 736--740, 2020.

\bibitem{wang2021jittering}
W.~Wang and W.~Zhang, ``Jittering effects analysis and beam training design for {UAV} millimeter wave communications,'' \emph{{IEEE} Trans. Wireless Commun.}, vol.~21, no.~5, pp. 3131--3146, 2021.

\bibitem{zhao2023integrated}
J.~Zhao, F.~Gao, W.~Jia, W.~Yuan, and W.~Jin, ``Integrated sensing and communications for {UAV} communications with jittering effect,'' \emph{{IEEE} Wireless Commun. Lett.}, 2023.

\bibitem{liSmartphoneLocalizationAlgorithm2013}
W.~W.-L. Li, R.~A. Iltis, and M.~Z. Win, ``A smartphone localization algorithm using {{RSSI}} and inertial sensor measurement fusion,'' in \emph{2013 {{IEEE Global Communications Conference}} ({{GLOBECOM}})}.\hskip 1em plus 0.5em minus 0.4em\relax {Atlanta, GA}: {IEEE}, Dec. 2013, pp. 3335--3340.

\bibitem{sidi1997spacecraft}
M.~J. Sidi, \emph{Spacecraft dynamics and control: a practical engineering approach}.\hskip 1em plus 0.5em minus 0.4em\relax Cambridge university press, 1997, vol.~7.

\bibitem{love2008overview}
D.~J. Love, R.~W. Heath, V.~K. Lau, D.~Gesbert, B.~D. Rao, and M.~Andrews, ``An overview of limited feedback in wireless communication systems,'' \emph{{IEEE} J. Sel. Areas Commun.}, vol.~26, no.~8, pp. 1341--1365, 2008.

\bibitem{liu2020radar}
F.~Liu, W.~Yuan, C.~Masouros, and J.~Yuan, ``Radar-assisted predictive beamforming for vehicular links: Communication served by sensing,'' \emph{{IEEE} Trans. Wireless Commun.}, vol.~19, no.~11, pp. 7704--7719, 2020.

\bibitem{abu2018error}
Z.~Abu-Shaban, X.~Zhou, T.~Abhayapala, G.~Seco-Granados, and H.~Wymeersch, ``Error bounds for uplink and downlink {3D} localization in {5G} millimeter wave systems,'' \emph{{IEEE} Trans. Wireless Commun.}, vol.~17, no.~8, pp. 4939--4954, 2018.

\bibitem{bar2001estimation}
Y.~Bar-Shalom, X.~R. Li, and T.~Kirubarajan, \emph{Estimation with applications to tracking and navigation: theory algorithms and software}.\hskip 1em plus 0.5em minus 0.4em\relax John Wiley \& Sons, 2001.

\bibitem{lefferts1982kalman}
E.~J. Lefferts, F.~L. Markley, and M.~D. Shuster, ``Kalman filtering for spacecraft attitude estimation,'' \emph{Journal of Guidance, control, and Dynamics}, vol.~5, no.~5, pp. 417--429, 1982.

\bibitem{prieto2016context}
J.~Prieto, S.~Mazuelas, and M.~Z. Win, ``Context-aided inertial navigation via belief condensation,'' \emph{{IEEE} Trans. Signal Process.}, vol.~64, no.~12, pp. 3250--3261, 2016.

\bibitem{kay1993fundamentals}
S.~M. Kay, \emph{Fundamentals of statistical signal processing: estimation theory}.\hskip 1em plus 0.5em minus 0.4em\relax Prentice-Hall, Inc., 1993.

\end{thebibliography}
\end{document}